\documentclass[twocolumn,showpacs,preprintnumbers,amsmath,amssymb,prb,superscriptaddress,aps,10pt]{revtex4-1}

\usepackage{graphicx}
\usepackage{float}
\usepackage{overpic}
\usepackage[usenames,dvipsnames]{color}
\usepackage{textcomp}

\usepackage{graphics}
\usepackage{graphicx}
\usepackage{epsfig,psfrag}

\newcommand{\comment}[1]{}

\begin{document}

\title{Full counting statistics of persistent current}
\author{A. Komnik}
\affiliation{Institut f\"ur Theoretische Physik,
Ruprecht-Karls-Universit\"at Heidelberg,\\
 Philosophenweg 19, D-69120 Heidelberg, Germany}
 \affiliation{Physikalisches Institut,
Albert-Ludwigs-Universit\"at Freiburg,\\
 Hermann-Herder-Str. 3, D-79104 Freiburg i. Br., Germany}
\author{G. W. Langhanke}
\affiliation{Institut f\"ur Theoretische Physik,
Ruprecht-Karls-Universit\"at Heidelberg,\\
 Philosophenweg 19, D-69120 Heidelberg, Germany}

\date{\today}

\begin{abstract}

We develop a method for calculation of charge transfer statistics of persistent current in nanostructures {\color{black} in terms of the cumulant generating function (CGF) of transferred charge}. We consider a simply connected one-dimensional system (a wire) with arbitrary interactions and develop a procedure for the calculation of the CGF of persistent currents when the wire is closed into a ring via a weak link. For the non-interacting system we derive a general formula in terms of the two-particle Green's functions. We show that, contrary to the conventional tunneling contacts, the resulting cumulant generating function has a doubled periodicity as a function of the counting field. We apply our general formula to short tight-binding chains and show that the resulting CGF perfectly reproduces the known evidence for the persistent current. Its second cumulant 
turns out to be maximal at the switching points and vanishes identically at zero temperature. {\color{black} Furthermore, we apply our formalism for a computation of the charge transfer statistics of genuinely interacting systems. First we} consider a ring with an embedded Anderson impurity and employing a self-energy approximation find an overall suppression of persistent current as well as of its noise. {\color{black} Finally, we compute the charge transfer statistics of a double quantum dot system in the deep Kondo limit using an exact analytical solution of the model at the Toulouse point. We analyze the behaviour of the resulting cumulants and compare them with those of a noninteracting double quantum dot system and find several pronounced differences, which can be traced back to interaction effects.}

\end{abstract}


\maketitle

\section{Introduction}

Persistent current (PC) in ring-shaped nanostructures is one of the most fascinating phenomena in mesoscopic physics.\cite{PC1,Imry} Despite enormous amount of work invested in its study there are still numerous aspects, which are yet not fully understood. Perhaps the most interesting are the issues of how electronic correlations affect the PC and whether it is subject to fluctuations in clean systems without disorder.\cite{LesovikoldPC,moskalets:982,Semenov1} A quite natural extension of the latter topic is the question about the full counting statistics (FCS) of persistent current, which, to the best of our knowledge, has not been addressed yet. 
FCS is an interesting and insightful quantity best described as a probability $P(Q)$ to transfer $Q$ charges through a constriction or a device during a (very long) \emph{\color{black} measuring time} ${\cal T}$.\cite{nazarov2009quantum} {\color{black} Since the seminal paper of W. Schottky it is known, that the shot noise, which is the second cumulant of the FCS, contains interesting information about the charge of current carriers.\cite{Schottky1918}. This idea was very important in such a breakthrough as the measurement of the fractional charge in quantum Hall edge states devices.\cite{dePicciotto,glattli} It turns out, however, that the charge of current carrying excitations can be much more precisely determined if one uses the third order correlation of the transport current instead of the shot noise.\cite{PhysRevB.70.115305,KomnikSaleur} This quantity can very conveniently be extracted from the cumulant generating function (CGF) of the FCS. CGF is also known to satisfy in many cases the Gallavotti-Cohen relation, which is a direct generalization of the conventional fluctuation-dissipation theorem.\cite{Esposito2009,KomnikSaleur2} Apart of that the cumulants of higher orders can help to reveal important details about interactions of the given system with the environment, thereby yielding important insights.\cite{PhysRevLett.91.196601,nazarov2009quantum}
Very recently the concept of the FCS has even been successfully employed in the field of ultracold quantum gases, as the counting statistics of Rydberg aggregates was measured, which has helped to identify multi-particle correlation effects.\cite{Rydbergs}
}

Here we report a progress along several lines. First of all we develop a general approach, which allows for the evaluation of PC of \emph{arbitrarily} shaped structure which is closed to a ring via single or several weak links, and which can be applied also to interacting systems. 
It turns out that for quadratic Hamiltonians one can even derive a closed formula for the FCS
using the Green's functions (GFs) of the original open structure. 
We generalize this method to the calculation of the FCS and using it we discuss the charge transfer statistics of noninteracting as well as interacting systems and find that in case of a wire with an Anderson impurity there is a suppression of PC as well as of its noise at least for not too strong interactions. 
{\color{black}
Furthermore, we discuss the FCS of PC in two different kinds of double quantum dot Aharonov-Bohm interferometer (ABI) setups: a noninteracting one and a one, both quantum dots of which are driven into the Kondo regime. We find fundamental differences in the cumulant generating functions which can be attributed to interaction effects and predict pronounced differences in the resulting measurable cumulants of the FCS.  

The paper is organized as follows. In the next section we explain the details of our method of the FCS computation. Section \ref{NonInt} is then devoted to applications of the developed technique to noninteracting systems. After that Section \ref{Kondo_ABI} contains applications to two different types of interacting systems: to a ring with an embedded Anderson impurity and to a double quantum dot ABI. Finally,  Section \ref{Conclusions} offers a summary of the results.  
}
\section{General approach}
\label{Generals}

The canonical definition of the PC is based on the observation that if the partition function $Z$ of a given system depends on the magnetic flux $\Phi$ piercing it, then the associated conjugate quantity is the finite charge current, so that (we set $e=\hbar=c=1$ throughout the paper)
\begin{equation} 
 I_{PC}= - \frac{\partial F}{\partial \Phi} \, ,
\end{equation}
where $F$ is the free energy of the system. 
\begin{figure}
\centering
\epsfig{file=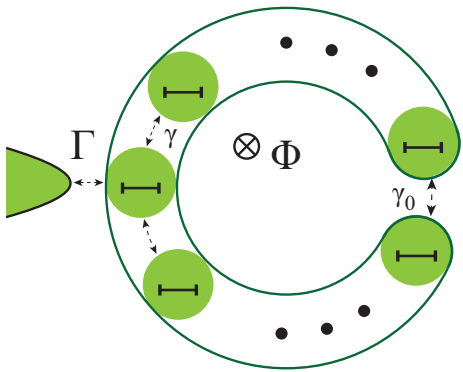,width=0.7\linewidth,clip=}
\caption{(Color online)
The setup of the system under consideration. For $\gamma_0=0$ it is simply connected. The wire can be coupled to a metallic electrode via hybridization $\Gamma$. 
\label{Fig1}}
\end{figure}
There are numerous ways of computing $I_{PC}$, ranging from quantum statistical to scattering methods
{\color{black} 
but most of them cannot yield the FCS of PC in a straightforward way. It turns out that the charge transfer statistics can rather easily be computed in systems which are characterized by one or several weak (tunnelling) links, at which one can introduce the counting field $\lambda$.\cite{PhysRevB.70.115305,PhysRevB.73.195301}  That is why a natural procedure is to take a ring, disconnect it and introduce at the `cut' point a weak link along with the counting field and compute the FCS just like it is done for transport in tunnelling systems.}

Therefore we would like to  
ask the following question: Let us 
suppose that we have a simply connected system, e.~g. a quantum wire with open boundaries, which we would like to shape into a ring via a weak link, connecting two arbitrary points of the wire. Can we derive an expression for PC using the GFs of the original open system? In order to answer this question 
one can proceed as follows. Let $H_0[\psi(x)]$ be the Hamiltonian describing the electronic degrees of freedom in an open simply connected wire (we concentrate on 1D wires, a generalization to higher-dimensional systems can be done along the same lines). It is closed to a ring by tunneling between the points $x=0$ and $x=L$ with the amplitude $\gamma_0$. Then the full system can be described by the following Hamiltonian,\footnote{We assume $\gamma_0$ to be purely real, which can always be done by appropriate gauge of $\varphi$.}
\begin{equation}
 H = H_0[\psi(x)]
 + \gamma_0 e^{-i \varphi} \, \psi^\dag(0) \, \psi(L) +  \gamma_0 e^{i \varphi} \, \psi^\dag(L) \, \psi(0) \, ,
\end{equation}
where $\varphi = 2 \pi \Phi/\Phi_0$ describes the magnetic flux enclosed by the wire measured in units of the flux quantum $\Phi_0 = h/e$. 
In the following we consider a situation 
when
the partition function can be written down as a functional integral over the \emph{local} fields $\phi_0(n)=\psi_n(0)$ and 
$\phi_L(n)=\psi_n(L)$, where $n$ denotes the respective Matsubara component, 
\begin{eqnarray}   \label{longZ}
 &&Z = Z_0  \,
    \int {\cal D} \left[\phi_0,\phi_L,\bar{\phi}_0,\bar{\phi}_L\right] \, \exp
   \bigg\{
     -  \frac{1}{\beta} \sum_n \left[
     \bar{\phi}_0(n) \, G_{00}^{-1} \,
     \right.  \nonumber \\ 
  &&\times \left.  \phi_0(n)   +  \bar{\phi}_L(n) \, G_{LL}^{-1} \, \phi_L(n)
     -  \bar{\phi}_L(n) \, \left( G_{L0}^{-1}  - \gamma_0 \, e^{i \varphi}\right) \, 
 \right.  \nonumber \\
    &&\times \phi_0(n) -  \left.  \bar{\phi}_0(n) \,\left( G_{0L}^{-1} - \gamma_0 \, e^{- i \varphi}\right) \, \phi_L(n) \right] \bigg\} \, ,
\end{eqnarray}
where $G_{00}$, $G_{LL}$, $G_{L0}$ and $G_{0L}$ are the corresponding two-particle Green's functions. In the case of a quadratic action they can easily be evaluated by integrating out 
all degrees of freedom away from $x=0,L$ using the standard methods. 
The GFs entering the above expression, can be shown to descend from the following two independent Matsubara GFs: 
(i) the \emph{local} one 
$G_0(n) = - \langle T \phi_x(n) \,\bar{\phi}_x(n) \rangle$
, which involves only fields at the contact points, in our case $x=0,L$ (from now on we assume both GFs to be identical due to spatial inversion symmetry of the open system \footnote{This is not a fundamental restriction. Similar formulas can be derived for arbitrary wire geometries.\label{foot1}});  
(ii) and the \emph{non-local} one connecting both contact points
 $G_L(n) = - \langle T \phi_L(n) \,\bar{\phi}_0(n) \rangle = - \langle T \phi_0(n)\, \bar{\phi}_L(n) \rangle$. 
 While the GFs {\color{black} of the type} (i) reflect the local density of states in the contact points of the weak link, the GF (ii) describes the single-particle propagation dynamics \emph{between} the contact points along the wire.
With the help of these definitions we find
\begin{eqnarray}  \label{Mats_1}
 G_{00} = G_{LL} &=& G_0 - G_L \, G_0^{-1} \, G_L \,  \nonumber\\
 G_{0L} =  G_{L0} &=& G_0 \, G_L^{-1} \left(G_0 - G_L \, G_0^{-1} \, G_L \right)  \nonumber\\
&=& G_{00} \, G_L^{-1} \, G_0 \, .
\end{eqnarray}
\comment{
\begin{eqnarray}  \label{Mats_1}
 G_{00}^{-1} &=& G_{LL}^{-1}= \left( G_0 - \widetilde{G}_L \, G_0^{-1} \, G_L \right)^{-1} \,  \nonumber 
 \\
 G_{L0}^{-1} &=&
  G_0^{-1} \, G_L \,  \left( G_0 - \widetilde{G}_L \, G_0^{-1} \, G_L \right)^{-1}
  \nonumber \\  \label{Mats_1}
 G_{0L}^{-1} &=& \left(G_0 - \widetilde{G}_L \, G_0^{-1} \, G_L \right)^{-1} \widetilde{G}_L \, G_0^{-1}  \, .
\end{eqnarray}
}
Using these relations we then immediately obtain the partition function for our composite system: 
\begin{eqnarray}           \label{nakedZ}  \nonumber
 \frac{Z_\varphi}{Z_0} =
  \prod_n \, \left( 1 - \gamma_0^2 \, G_{00} \, G_0 + 2\gamma_0 \, G_{L} \cos \varphi \right) \, ,
\end{eqnarray}
\comment{
\begin{eqnarray}           \label{nakedZ}  \nonumber
 \frac{Z_\varphi}{Z_0} =
  \prod_n \, \left[ 1 - \gamma_0^2 \, G_{LL} \, G_0 \gamma_0 \,  \left( G_{L} e^{- i \varphi}  +
    \widetilde{G}_{L} e^{ i \varphi} \right) \right] 
    \, ,
\end{eqnarray}
}
where $Z_0$ is the partition function of the open system ($\gamma_0 =0$).
The emerging PC is then given by 
\begin{eqnarray} \label{classic} \nonumber
  I_{\rm PC} = \frac{1}{\beta} \sum_n \frac{2\gamma_0 \,G_{L} \sin \varphi }{  1 - \gamma_0^2 \, G_{00} \, G_0 + 2 \gamma_0 \,{G}_{L} \cos \varphi } \, .
\end{eqnarray}
\comment{
\begin{eqnarray} \label{classic} \nonumber
  I_{\rm PC} = - \frac{1}{\beta} \sum_n \frac{i \gamma_0  \left( G_{L} e^{i \varphi} - \widetilde{G}_{L} e^{-i \varphi}\right)}{  1 - \gamma_0^2 \, G_{LL} \, G_0 + \gamma_0  \left( G_{L} e^{i \varphi} + \widetilde{G}_{L} e^{-i \varphi}\right) } \, .
\end{eqnarray}
}
With the help of this relation one can recover all known results for the PC in non-interacting systems, see e.~g. Ref.~[\onlinecite{PhysRevB.50.4921}].

Now we would like to extend our formalism to the computation of the FCS in terms of the cumulant generating function (CGF) $\ln \chi(\lambda)$ of $P(Q)$.\cite{nazarov2009quantum} There are different ways to obtain it, the procedure we want to implement is the one of Ref.~[\onlinecite{PhysRevB.70.115305}]. It consists of multiplying every forward/backward tunneling term in the Hamiltonian by the factor $e^{\pm i \lambda/2}$, where $\lambda$ is the counting field, which carries opposite sign on the forward/backward Keldysh branch, see e.~g. Ref.~[\onlinecite{PhysRevB.73.195301}]. In this particular procedure the charges are counted only during some very long measuring time ${\cal T}$. Before and after that time span the system is
\emph{disconnected}. So the advantage offered by our method is that everything can be represented in terms of GFs of an open system, which, being simply connected, underlies simpler boundary conditions. The necessity for employment of Keldysh techniques roots in the clear distinction of system states when the tunneling is switched on and when it is switched off. Gell-Mann and Low theorem does not hold and one has to resort to non-equilibrium techniques.\cite{mahan}

There is, however, one decisive detail, which makes our ring-shaped setup completely different from the orthodox tunneling systems, for which the FCS procedure was designed. In tunneling systems one usually considers the scattered particles as coming from and vanishing into the `infinity' -- the incoherent background of the electrodes. At least from the ideological point of view this is one of the reasons the counting procedure works, all tunneling events are distinguishable from each other. This is not so for a ring. The particles which are already counted once are coming back in coherent fashion, and there is a possibility that the whole procedure would not give meaningful results. This problem can, however, be circumvented by introduction of an additional very large particle bath -- for instance just an additional electrode as is done in e.~g. Refs.~[\onlinecite{But1}] and [\onlinecite{But2}]. In this way the electron states in the ring are hybridized with the continua in the electrode and so become delocalised and reach to `infinity' just like the scattering states in a conventional tunneling junction. Thus, from the point of view of the junction, where
the charges are counted, both a contacted ring and an open tunneling
junction are indistinguishable. That is why our approach works. We shall see later though, that the presence of the electrode is not crucial and that the 
conventional FCS method works well even for very short tight-binding chains.
Nonetheless, sometimes it is useful to keep the extra electrode as in some experiments it is explicitly present and used to measure the PC.\cite{Kang} {\color{black} Of course, it is essential to make sure that the resulting CGF recovers all already known facts about PC. Therefore, before applying our method to interacting systems, in what follows we make explicit calculations for simple noninteracting rings and show that our technique yields exactly the same results as all other methods used previously.}

The FCS computation is easiest in the Keldysh representation, in which all fields have two components $\boldsymbol{\psi}(x) = (\psi_-(x), \psi_+(x))$, describing the forward/backward ($\pm$) time propagation. All GFs have then 4 different components, 
\begin{eqnarray}
 {\bf G}(x-x',t-t') = \left(
 \begin{array}{cc}
 G^{--} & G^{-+} \\
 G^{+-} & G^{++}
 \end{array}
 \right) \, ,
\end{eqnarray}
where $G^{--}$ and $G^{++}$ are the time-ordered and anti-time-ordered components, and $G^{-+}$ and $G^{+-}$ are the lesser and greater Keldysh GFs.\cite{lifshitz1981physical} Using this language and assuming a quadratic action we can again integrate out all fields away from the contact points and end up with the following generalization of the partition function \eqref{longZ}:
\begin{widetext}
\begin{eqnarray}
  \label{big_Z}
  Z = Z_0 \,
    \int {\cal D} \left[\boldsymbol{\phi}_0,\boldsymbol{\phi}_L,\bar{\boldsymbol{\phi}}_0, \bar{\boldsymbol{\phi}}_L\right] \, \exp
   \bigg\{
     &-&  \int \frac{d \omega}{2 \pi}  \left[
     \boldsymbol{\bar\phi}_0(\omega) \, {\bf G}_{00}^{-1} \, 
  \boldsymbol{\phi}_0(\omega)
     +  \boldsymbol{\bar\phi}_L(\omega) \, {\bf G}_{LL}^{-1} \, \boldsymbol{\phi}_L(\omega)
     -  \boldsymbol{\bar\phi}_L(\omega) \, \left[{\bf G}_{L0}^{-1} -   \boldsymbol{\gamma}_0\right]\, 
     \right. \nonumber \\
   &&\times \left.   \boldsymbol{\phi}_0(\omega)
     -  \boldsymbol{\bar\phi}_0(\omega) \, \left[{\bf G}_{0L}^{-1}  -  \boldsymbol{\gamma}_0^* \right] \, \boldsymbol{\phi}_L(\omega) \right] 
     \bigg\} \, ,
\end{eqnarray}
\end{widetext}
where the counting field enters the matrices
\begin{eqnarray}    \label{lambda}
  \boldsymbol{\gamma}_0 = \left( \begin{array}{cc}
 \gamma_0 \,e^{i (\lambda/2+\varphi)} & 0 \\
 0 & -\gamma_0 \,e^{-i (\lambda/2 - \varphi)}
 \end{array} \right) \, .
\end{eqnarray}
The analogs of \eqref{Mats_1} are now given by
\begin{eqnarray}
  {\bf G}_{00} = {\bf G}_{LL} &=& {\bf G}_0 - {\bf G}_L \, {\bf G}_0^{-1} \, {\bf G}_L \nonumber \\ 
  {\bf G}_{0L} = {\bf G}_{L0} &=&   {\bf G}_{00} \, {\bf G}_{L}^{-1} \, {\bf G}_{0}  \, .
\end{eqnarray}
\comment{
\begin{eqnarray}
 {\bf A} &=& \sigma_z \, {\bf G}_0 \, \sigma_z -
 \sigma_z \, \widetilde{\bf G}_L \,
 {\bf G}_0^{-1} \, {\bf G}_L \, \sigma_z \nonumber \\ 
 {\bf G}_{00}^{-1} &=&  \sigma_z \, {\bf A}^{-1} \,  \sigma_z
  \nonumber \\ 
 {\bf G}_{LL}^{-1} &=& {\bf G}_0^{-1} + 
 {\bf G}_0^{-1} 
   \, {\bf G}_L \,  \sigma_z \, 
    {\bf A}^{-1} \, 
    \sigma_z \, \widetilde{\bf G}_L  \, {\bf G}_0^{-1} 
    =  {\bf G}_{00}^{-1} \nonumber \\ 
 {\bf G}_{L0}^{-1} &=&
 {\bf G}_0^{-1}  
    \, {\bf G}_L \,  \sigma_z \, {\bf A}^{-1} \,  \sigma_z 
    \nonumber \\ 
 {\bf G}_{0L}^{-1} &=&  \sigma_z \, {\bf A}^{-1} \,  \sigma_z 
 \widetilde{\bf G}_L \,  {\bf G}_0^{-1} =  {\bf G}_{L0}^{-1}  \, ,
\end{eqnarray}
}
The CGF is up to a prefactor given by the $\lambda$-dependent Keldysh partition function of the system, therefore performing the last remaining functional integrations in \eqref{big_Z} we arrive at our principal result,
\begin{eqnarray}   \label{princ_Res}
 \ln \chi(\lambda) &=& 
{\cal T} \int d \omega \, \ln \, \mbox{det} \, \boldsymbol{\Lambda} \, ,
   \\ \nonumber
 \boldsymbol{\Lambda} &=& {\bf G}_{00}^{-1}     
        -  ({\bf G}_{0L}^{-1}  -  \boldsymbol{\gamma}_0^* ) {\bf G}_{00} 
         ({\bf G}_{0L}^{-1} -   \boldsymbol{\gamma}_0) \, ,
\end{eqnarray}
\comment{
\begin{eqnarray}   \label{princ_Res}
 \ln \chi(\lambda) &=& G_0 {\cal T} \int d \omega \, \ln \, \mbox{det} \, \boldsymbol{\Lambda} \, ,
   \\ \nonumber
 \boldsymbol{\Lambda} &=& {\bf G}_{00}^{-1}     
        -  ({\bf G}_{0L}^{-1}  -  \boldsymbol{\gamma}_0^* ) {\bf G}_{LL} 
         ({\bf G}_{L0}^{-1} -   \boldsymbol{\gamma}_0) \, ,
\end{eqnarray}
}
where 
${\cal T}$ is the measuring time.
It is instructive to compare it with the statistics of a simple structureless tunneling contact with the Hamiltonian
\begin{eqnarray} \nonumber
 H = H_0[\psi_{1,2}(x)] + \gamma_0 \psi_1^\dag(0) \, \psi_2(0) +  \gamma^*_0  \, \psi_2^\dag(0) \, \psi_1(0) \, ,
\end{eqnarray}
with $\psi_{1,2}(x)$ describing the electron fields of the right/left contact. The corresponding FCS is given by \cite{levitov-1996-37,PhysRevB.73.195301}
\begin{eqnarray}  \label{tun_con}
 \ln \chi(\lambda) = 
{\cal T}  \int d \omega \, \ln \, \mbox{det}
  \left( {\bf G}_{1}^{-1} 
        - \boldsymbol{\gamma}_0^* \, {\bf G}_{2} \,
         \boldsymbol{\gamma}_0  \right) \, ,
\end{eqnarray}
where ${\bf G}_i$ denotes the local GF on the respective electrode at the tunneling point. 
As ${\bf G}_{00}$ 
in the wire geometry also describes the strictly local GF we have an immediate parallel ${\bf G}_{00}\leftrightarrow {\bf G}_{1}$ \emph{and} ${\bf G}_{00} = {\bf G}_{LL} \leftrightarrow {\bf G}_{2}$. 
Therefore we can split the object  $\boldsymbol{\Lambda} =
 \boldsymbol{\Lambda}_{\rm c} +  \boldsymbol{\Lambda}_{\rm ic}$ into the \emph{incoherent} contribution which has the same shape as the matrix for the tunneling contact,
 \begin{eqnarray}
  \boldsymbol{\Lambda}_{\rm ic} =
  {\bf G}_{00}^{-1}     
        -   \boldsymbol{\gamma}_0^* \, {\bf G}_{00} 
        \, \boldsymbol{\gamma}_0 \, ,
 \end{eqnarray}
and the \emph{coherent} part
\begin{eqnarray}
   \boldsymbol{\Lambda}_{\rm c} =     
         \boldsymbol{\gamma}_0^*  {\bf G}_{00} 
         {\bf G}_{0L}^{-1} 
          +  {\bf G}_{0L}^{-1}  {\bf G}_{00} 
            \boldsymbol{\gamma}_0 \, ,
\end{eqnarray}
which is absent in the tunneling contact case and which reflects the fact that in the case of the doubly connected system both contacts `talk' to each other through the wire itself. 
Its presence has interesting consequences. 
While the conventional FCS of noninteracting systems turns out to be $2\pi$-periodic in $\lambda$, the FCS of PC has a doubled periodicity.
This can be understood in the following way. 
The term $\boldsymbol{\Lambda}_{\rm ic}$ is of the order $\gamma_0^2$ and arises from the tunneling of the electron across the weak link and back (forward and backward propagation along the Keldysh contour). On the contrary, the coherent term $\boldsymbol{\Lambda}_{\rm c}$ is linear in $\gamma_0$. That means that the counted electron tunnels through the weak link on the forward (backward) path and returns back following the wire rather than tunneling back directly. {\color{black} On the other hand, this periodicity doubling can also be interpreted in terms of counting particles in a tunnelling system, which is initially prepared in a superposition of different charge states, see e.~g. Refs.~[\onlinecite{Shelankov}] and [\onlinecite{LesovikSadovskii}]. }

\section{Applications to noninteracting systems}
\label{NonInt}

\begin{figure}
\centering
\epsfig{file=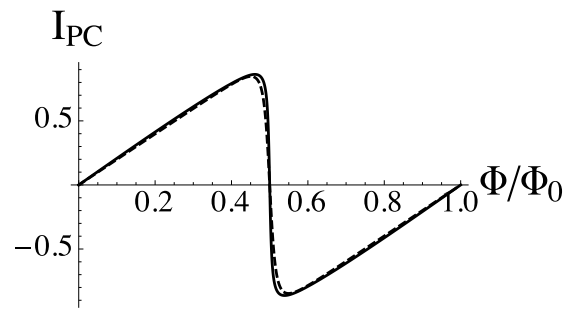,width=0.85\linewidth,clip=}
\caption{
Persistent current through a ring with the length $N=3$ computed with the help of our approach using the FCS of the PC (solid line) and the formula (1) given in Ref.~[\onlinecite{PhysRevB.50.4921}] (dashed line) at zero temperature, measured in units of $\gamma/2 \Phi_0$. $\gamma$ is the hybridization amplitude between the adjacent sites within the chain, the tunnelling amplitude of the weak link is $\gamma_0/\gamma=0.9$, which corresponds to $T_F\approx 0.989$ as it appears in Ref.~[\onlinecite{PhysRevB.50.4921}]. The chemical potential is set to zero and $\Gamma/\gamma=0.025$. The agreement of both curves improves for descreasing $\Gamma$.
\label{FigComparison}}
\end{figure}

The individual cumulants of the charge transfer statistics are computed as usual using the prescription 
\begin{eqnarray}
 C_n = (-i)^n \frac{\partial^n  \ln \chi(\lambda)}{\partial \lambda^n}\Big|_{\lambda \to 0} \, .
\end{eqnarray}
Thus $I_{\rm PC} = C_1/{\cal T}$. We have tested the prediction for the PC found using  \eqref{princ_Res} and \eqref{classic} for a tight-binding chain with $N$ sites (we take odd $N$ in order so satisfy the symmetry requirements imposed on the GFs), connected by hopping integral $\gamma$, which sets the energy scale of the system, and fixed chemical potential. We have compared it with the values for $I_{\rm PC}$ for the continuum model of finite length given in 
Ref.~[\onlinecite{PhysRevB.50.4921}]. 
In one case we have coupled the central chain site to an additional metallic electrode (bath) with some small hybridization $\Gamma$.
We find an excellent agreement already for very short chains with $N=3$ as long as the hybridization is the smallest energy scale, see Fig.~\ref{FigComparison}.\cite{DiplomGerald} Furthermore we considered a fully decoupled system in which $\Gamma=0$, obtaining precisely the same results as the classical formulas and recovering all important details such as the parity effect and the $1/N$ dependence of the PC on the chain length $N$ and exponential suppression of $I_{\rm PC}$ with growing temperature.\cite{DiplomGerald,longPC} {\color{black} Moreover, we have computed the two lowest order cumulants directly using the current autocorrelation function and found exactly the same results.}

\begin{figure}
\centering
\epsfig{file=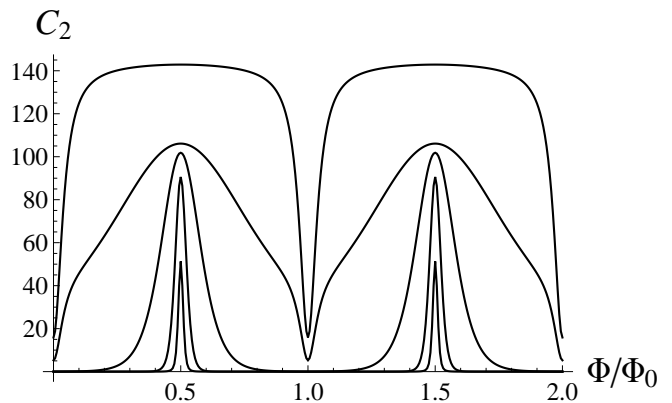,width=1.\linewidth,clip=}
\caption{
Second cumulant $C_2$ measured in units of $e \gamma {\cal T}/2\Phi_0$
as a function of magnetic flux for different values of temperature, from above $T/\gamma = 1.6, 0.4, 0.1, 0.025, 0.00625$. The energy level width is kept at $\delta/\gamma=0.025$ and $\gamma_0/\gamma=0.9$. 
\label{Fig3}}
\end{figure}

In Fig.~\ref{Fig3} we plot the second cumulant computed for a decoupled 3-site tight-binding chain with finite very small energy level widening $\delta$. 
The generic feature is the rapid decay of $C_2$ towards lower temperatures until it ultimately vanishes at $T=0$. This is to be expected since at zero temperature the system is in its ground state and all fluctuations are frozen out.   
This finding is compatible to the results of Refs.~[\onlinecite{Semenov1}],[\onlinecite{PhysRevB.84.045416}]. 
Another important feature is the maximum located precisely at half period.\cite{moskalets:982} 
This is not surprising either since at precisely these points the PC changes sign, therefore the probability for current fluctuations is the highest. 
$C_2$ shows an interesting behavior as a function of temperature. 
In the case of tight-binding chains we find $C_2(T) \sim \alpha_1 - \alpha_2/[1 + (T/T^*)^2]$, where $\alpha_{1,2}$ are some model specific constants and $T^*$ is the energy scale set by $E_F$. It turns out to be very close to the characteristic temperature on which the exponential suppression of PC itself occurs. 

\begin{figure}
\centering
\epsfig{file=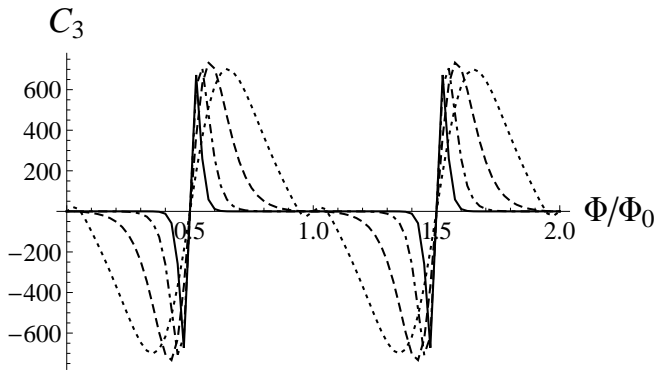,width=1.\linewidth,clip=}
\caption{
Third cumulant measured in units of $e^2 \gamma {\cal T}/4\Phi_0$
at different temperatures $T/\gamma = 0.25, 0.125, 0.0625, 0.03125$ (dotted, dashed, dashed-dotted and solid lines, respectively) for $\gamma_0/\gamma=0.9$ in a coupled system with $\Gamma/\gamma=0.05$. 
\label{thirdCFig}}
\end{figure}

Contrary to the second cumulant the third one does not vanish at zero temperature at least in the vicinity of the turning point $\Phi/\Phi_0=1/2 \text{ or } \varphi=\pi$, see Fig.~\ref{thirdCFig} and turns out to have a singularity there. One interesting peculiarity is also the very strong dependence of the third cumulant on the coupling to the lead $\Gamma$, the asymptotic value of $C_3$ at $\Phi/\Phi_0=1/2$ approximately growing as $\sim 1/\Gamma$. While the strong dependence of the third cumulant on the effects of the environment has been found before even experimentally,\cite{PhysRevLett.91.196601} it is not clear yet whether this singularity  can be explained by the same physical mechanism. 
\begin{figure}
\centering
\epsfig{file=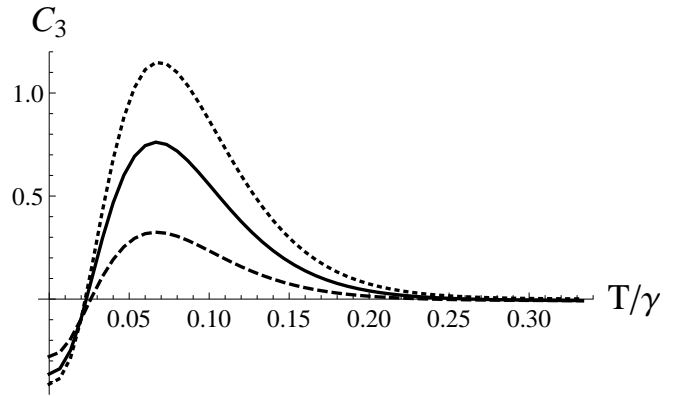,width=1.\linewidth,clip=}
\caption{
The temperature dependence of the third cumulant measured in units of $e^2 \gamma {\cal T}/4\Phi_0$
$\gamma_0/\gamma=1, 0.7, 0.5$ (dotted, solid and dashed line, respectively) and $\varphi=\pi/4$ or $\Phi/\Phi_0 = 1/8$ for a tight-binding chain with $N=15$ and zero chemical potential. 
\label{thirdCTempDep}}
\end{figure}
$C_3$ shows a non-monotonous behavior as a function of temperature, see Fig.~\ref{thirdCTempDep}. It has a distinct maximum at intermediate temperatures and decays exponentially towards the limit $T/\gamma \to \infty$. This is very similar to the third cumulant for a tunneling contact between two non-interacting metals biased by finite voltage with an \emph{energy-dependent} transmission, vanishing in the large energy limit. This is remarkable because a simple weak-link is known to have an energy-independent transmission coefficient.

Another interesting feature of $C_3$ is its exponential decay with the length of the system. A detailed study of these properties as well as of the higher order cumulants will be presented elsewhere.\cite{longPC} 
The general tendency is that at zero temperature all even order cumulants vanish and all odd order objects show a behavior similar to that of the third one.

\begin{figure}
\centering
\begin{minipage}{0.45\linewidth}
\epsfig{file=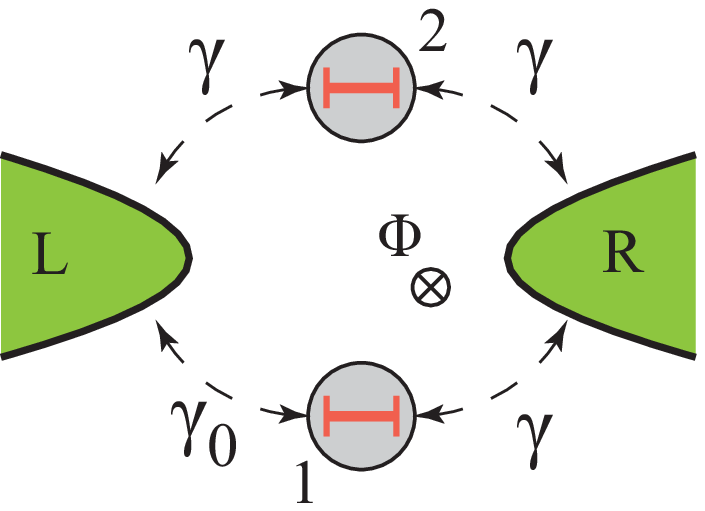,width=1.\linewidth,clip=}
\end{minipage}
\begin{minipage}{0.5\linewidth}
\epsfig{file=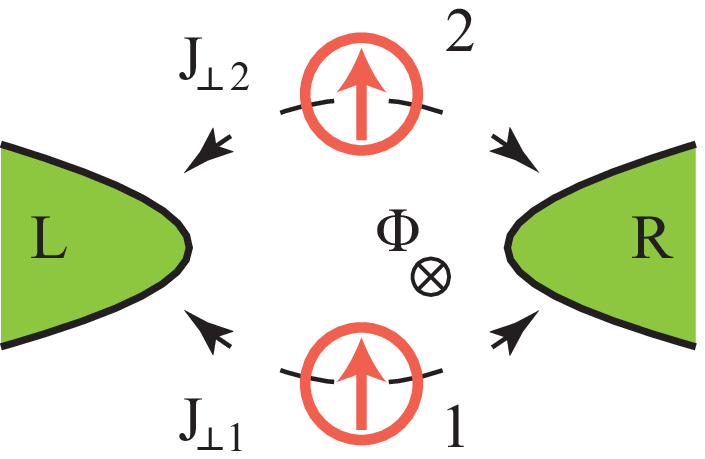,width=1.\linewidth,clip=}
\end{minipage}
\caption{ \color{black}(Color online)
 \emph{Left panel:}
 A double quantum dot Aharonov-Bohm interferometer formed by two metallic electrodes $L$, $R$ and two quantum dots $1$ and $2$. 
 \emph{Right panel:} A double quantum dot Aharonov-Bohm interferometer in the Kondo limit. The transport between the electric leads is possible via spin-flip processes with amplitudes $J_{\perp 1,2}$. 
 \label{AB_Toulouse}}
\end{figure}
{\color{black}
As another application we would like to analyze the FCS of a double quantum dot ABI, see Fig.~\ref{AB_Toulouse}. This is a versatile device widely used to study transport at nanoscale.\cite{DDreview} We would like to consider its simplest realization, consisting of two metallic electrodes, the electrons in which are described by fermion fields $R(x)$ and $L(x)$ and two localized fermionic levels with energies $\Delta_{i}$ ($i=1,2$), the respective electron creators/annihilators being denoted by $d^\dag_i$, $d_i$. The corresponding Hamiltonian is given by:
\begin{eqnarray}
H=H_1 + H_t \, , \, \, \, \, \, 
 H_1 = H_0[R,L] + \sum_{i=1,2} \Delta_i \, d_i^\dag d_i^\dag \, ,
\end{eqnarray}
where the electrodes are coupled to both localized levels by tunnelling amplitudes $\gamma$ and $\gamma_0$:
\begin{eqnarray}
 H_t &=&  \gamma [R^\dag d_1 
  + d_1^\dag R + d_2^\dag (R + L) + (R^\dag + L^\dag) d_2]
  \nonumber \\
  &+& \gamma_0 (e^{i \varphi} d_1^\dag L +  e^{-i \varphi}  L^\dag d_1 ) \, ,
\end{eqnarray} 
where the tunnelling from/into the electrodes is assumed to occur locally at $x=0$ in the coordinate system of the respective electrode, so that $R=R(x=0)$ and $L=L(x=0)$. We shall describe the electrode degree of freedom in the framework of the wide flat band model (WFBM) for simplicity, but this is not a severe restriction and can easily be relaxed. 

Such kind of system is an unorthodox realization of a ring geometry one usually considers in the theory of PCs because of the coupling to \emph{two} different electrodes. We shall construct a formal solution for the system in the general case of finite bias voltage applied between the electrodes and discuss in detail only the zero voltage case. Furthermore, we shall  consider only such quantities, which are only present when there is a finite PC in a system. The reason for that is the fact that for instance the second cumulant, or the noise, can be finite even at zero voltage at finite temperature but no field $\varphi=0$, thereby yielding information which cannot be connected to the PC phenomenon at all. It is obvious, that along with the second cumulant one must exclude all even order cumulants. Odd order cumulants, on the contrary, at least at zero temperature are finite only in finite $\varphi \neq 0$.  Therefore we concentrate on the PC and the third cumulant of current $C_3$ only. Moreover, contrary to the situations considered above we would like not to make the tunnelling amplitudes to the electrodes small. We believe that such a geometry can be quite advantegeous for future experimental investigations of PCs. 

In order to compute the CGF we first have to build the counting field into the system Hamiltonian. In the present case it is most efficiently done by a transformation
\begin{eqnarray}  \nonumber
 e^{i \varphi} d_1^\dag L +  e^{-i \varphi}  L^\dag d_1 \to
 e^{i (\varphi+\lambda/2)} d_1^\dag L +  e^{-i (\varphi+\lambda/2)}  L^\dag d_1 \, .
\end{eqnarray}
We would like to point out again, that $\lambda$ is a field which has opposite sign on the forward/backward paths of the Keldysh contour (as opposed to $\varphi$). The CGF is now nothing else but the $\lambda$-dependent Keldysh partition function. It can either be computed by differentiation with respect to the coupling constant, or by a straightforward functional integration.\cite{PhysRevB.73.195301} We follow the second route and define the full action of the system in form of a matrix: 
\begin{eqnarray}      \label{BigT}
 A = \left( 
\begin{array}{cccc}
 {\bf G}_L^{-1} & {\bf 0} & {\bf T}_\lambda &  {\bf T}_0 \\
 {\bf 0} &  {\bf G}_R^{-1} & {\bf T}_0 & {\bf T}_0 \\
 {\bf T}_\lambda^* & {\bf T}_0^* &  {\bf D}_1^{-1} & {\bf 0} \\
  {\bf T}_0^* &  {\bf T}_0^* & {\bf 0} & {\bf D}_2^{-1}
\end{array} \right) \, ,
\end{eqnarray}
where $ {\bf G}_{L,R}^{-1}$ are the local Keldysh GFs for the electrons in the respective electrodes with the structure $G_{s}(t) = - i \langle T_C s(x=0,t) s^\dag(x=0,0) \rangle$ with $s=R,L$ and $T_C$ being the contour ordering operator. Within the WFBM they are very simple and given by
\begin{eqnarray}
 {\bf G}_s(\omega) =  i \rho_0  \left( \begin{array}{cc}
 n_s- 1/2 &  n_s \\
 n_s -1 &  n_s -1/2 \\
 \end{array}\right) 
\end{eqnarray}
with $\rho_0$ being the density of states in the electrode and $n_s$ being the Fermi distribution function in the respective electrode. Keldysh GFs of the uncoupled dots are quite simple as well: ${\bf D}_i=\mbox{diag} (1/(\omega - \Delta_i), -1/(\omega - \Delta_i))$. Finally, the tunnelling contributions in the action are diagonal in the Keldysh space since the tunnelling processes are assumed to be instantaneous and can be shown to be given by ${\bf T}_0 = \mbox{diag} (\gamma, - \gamma)$ and 
 ${\bf T}_\lambda =  \boldsymbol{\gamma}_0$, see Eq.~\eqref{lambda}.
The CGF is then a functional integral of the kind
\begin{eqnarray}
 \ln \chi(\lambda) &=& \ln \int {\cal D}[\boldsymbol{\alpha}^\dag, \boldsymbol{\alpha}] \, \exp\left[- \int d \omega \, \boldsymbol{\alpha}^\dag \, A \boldsymbol{\alpha}\right]
 \nonumber \\
 &=& {\cal T} \int d \omega \, \ln \det A \, ,
\end{eqnarray}
where $\boldsymbol{\alpha} = (R_-, R_+, L_-, L_+, d_{1-}, d_{1,+}, d_{2,-}, d_{2,+})^T$ is the superfield used to construct the action. The subscript $\pm$ denotes the Keldysh branch index of the respective field.  

Although the CGF itself is quite involved the PC in the symmetric case $\gamma_0=\gamma$ and $\Delta_{1,2}=\Delta$ is given by the following compact relation:
\begin{eqnarray}      \label{PC_DD}
 I_{PC} &=& 8 \frac{\Gamma^3}{\Phi_0}  \sin \varphi  \int d \omega \, {(\omega - \Delta) (2n_F - 1)   }
\\  \nonumber &\times& 
 \left[16 \Gamma^4 \sin^4(\varphi/2) + 2 \Gamma^2 (\omega- \Delta)^2 ( 6 + 2 \cos \varphi)
 \right. \nonumber \\ \nonumber
  &+& \left. (\omega - \Delta)^4\right]^{-1} \, , 
\end{eqnarray}
where $\Gamma = \pi \rho_0 \gamma^2$. Expressions for all other cumulants can also be written down in a closed form. However, they are rather lengthy and therefore we do not report them.

The behaviour of the PC is generic and is shown in Fig.~\ref{Nonint_DD_IPC_C3}. As can be clearly seen already from \eqref{PC_DD} it is maximal when the system is in resonant configuration with $\Delta_{1,2}=0$. 
\begin{figure}[htbp]
\centering
\includegraphics[width=0.5\textwidth]{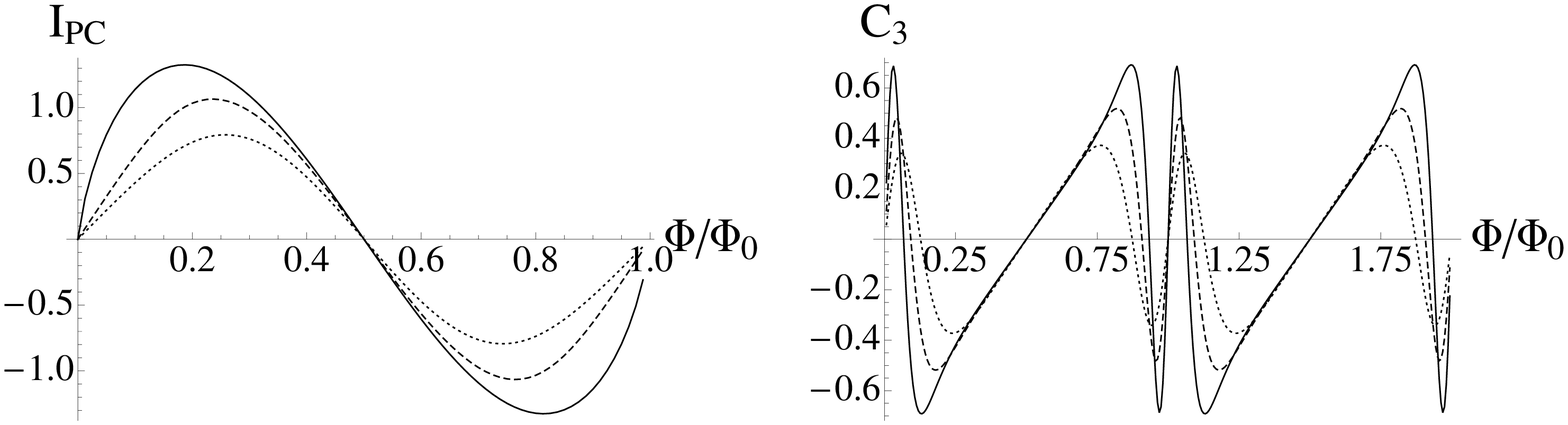}
\caption{\color{black}
\emph{Left panel:} Zero temperature PC through a noninteracting double dot ABI setup, measured in units of $\Gamma/\Phi_0$ for different dot offset energies $\Delta/\Gamma = 0$, $0.5$, $1.0$ (solid, dashed, dotted lines) and $\gamma_0=\gamma$.  
\emph{Right panel:} the third cumulant of the persistent current measured in units of $e^2 \Gamma {\cal T}/4\Phi_0$ for the same system at resonance $\Delta_{1,2}=0$ for different temperatures: $T/\Gamma=0.05$, $0.1$, $0.2$ (solid, dashed and dotted lines).
\label{Nonint_DD_IPC_C3}}
\end{figure}
The behaviour of the third cumulant qualitatively closely resembles the behaviour of $C_3$ for rings with a single electrode, see Fig.~\ref{Nonint_DD_IPC_C3}. It is nonzero even at zero temperature and at resonance $\Delta_{1,2}=\Delta$ has a singularity around a turning point, which is very similar to the one observed in the case of ordinary rings, see Fig.~\ref{thirdCFig}.  However, it is cut off at finite temperatures or/and finite gate voltage $\Delta_{1,2}\neq 0$. 

It is very well known from the theory of double quantum dots, that the transmission coefficient for the particles travelling between the electrodes is significantly suppressed for the parameter set $\Delta_1 = -\Delta_2= \Delta$.\cite{PhysRevB.68.125326,PhysRevB.65.245301,Dahlhaus} It turns out, that it is different for the PC. It is highest for $\Delta_{1,2}=0$ as expected. However, the symmetric gating $\Delta_1=\Delta_2$ leads to considerably smaller PC and $C_3$ as in the situation with $\Delta_1 = -\Delta_2$.
}

\section{FCS of the persistent current in interacting systems}
\label{Kondo_ABI}

\begin{figure}
\centering
\epsfig{file=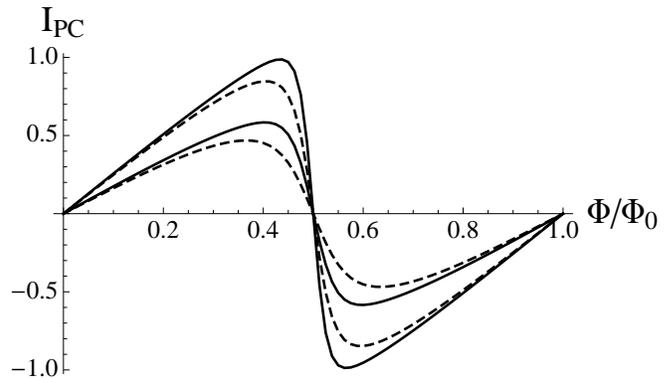,width=1.\linewidth,clip=}
\caption{
A comparison between the PC in a non-interacting system (solid lines) with $\Gamma/\gamma=0.05$, $\gamma_0/\gamma=0.9, 0.45$ (upper/lower curves) and an interacting system with $U/\gamma=0.4$ (dashed lines with the same ratios $\gamma_0/\gamma$) {\color{black} at temperature $T/\gamma=0.02$}. We have deliberately used a too high value of $U$ in order to demonstrate the relative weakness of the interaction effects {\color{black} in this approximation}. The current is measured in units of $\gamma/2\Phi_0$. {\color{black} The corresponding Kondo temperature is $T_K/\gamma \approx 0.012$.}
\label{Fig4}}
\end{figure}

Now we would like to turn to interacting systems. Although it is not always possible to find a representation of the form \eqref{big_Z}, the general procedure of closing an open system via term \eqref{lambda} and subsequent evaluation of the generalized $\lambda$-dependent Keldysh partition function is still perfectly applicable. To illustrate that we consider a situation in which 
one of the chain sites is replaced by an Anderson impurity -- in our case the role of the impurity plays the site, which is coupled to the electrode, see Fig.~\ref{Fig1}. Without interaction its energy is resonant (zero). Although the corresponding exact GFs are not known, there are plenty of powerful approximative techniques. In this introductory study we restrict ourselves to the approach using the second order self-energies,\cite{yamada2004electron} closed analytical expressions for which are e.~g. presented in Ref.~[\onlinecite{PhysRevB.83.075107}]. Fig.~\ref{Fig4} shows the results for the PC. For growing $U$ the PC as well as the noise turn out to be suppressed while keeping their overall non-interacting shape as a function of the magnetic flux. The decrease of PC is in agreement with general expectation and can be understood in terms of the decreasing impurity density of states at the Fermi edge due to the formation of the Hubbard sidebands. {\color{black} The temperature for the plot is chosen in such a way that it is almost twice as large as the Kondo temperature $T_K \approx \sqrt{U \Gamma} \exp[-\pi U/(8\Gamma)]$ below which a profound change in ground state occurs.\cite{Tsvelick1983} 
}

{\color{black}
At temperatures $T< T_K$ the Kondo effect dominates the low energy properties of the above system and the self-energy approximation is not applicable any more. While for the standard Kondo impurity problem there are a number of different solution approaches, to do so in a system with a ring geometry is a daunting problem, see e.~g. Refs.~[\onlinecite{Kondo_PC_Kang, Kondo_PC_Argentina,Kondo_PC_Affleck}]. However, in the impurity configuration proposed above the Kondo ground state is formed due to coupling to the external lead. This circumstance allows for a construction of a two-electrode Aharonov-Bohm setup with two imbedded Kondo impurities,   which possesses an exactly solvable Toulouse limit.\cite{PhysRevLett.87.156803,PhysRevLett.86.5128} 

The Hamiltonian of the system, see also the right panel of Fig.~\ref{AB_Toulouse}, is given by
\begin{eqnarray}\label{Hkondo_X}
 H = H_0 + H_J  \, ,
\end{eqnarray}
where, with $\psi_{\alpha, \sigma}$ being the electron
field operators in the $R,L$ electrodes,
\begin{eqnarray}\label{Hkondo}
 H_0 &=& i v_F \sum_{\alpha=R,L} \sum_{\sigma=\uparrow,\downarrow}
 \int \, dx \,
 \psi^\dag_{\alpha \sigma}(x) \partial_x \psi_{\alpha \sigma}(x) 
 + \sum_{i=1,2} \Delta_i \tau_i^z \, ,
 \nonumber \\
 H_J &=& \sum_{\alpha, \beta = R,L} \sum_{\nu=x,y,z} \sum_i
 J_{i \nu}^{\alpha \beta}
 s^\nu_{\alpha \beta} \sum_i \tau_i^\nu \, .
 \end{eqnarray}
Here $\tau_i^{\nu=x,y,z}$ are the Pauli matrices for the
impurity spins located at the dots $i=1,2$ and
\[
s^\nu_{\alpha \beta}=\sum_{\sigma,\sigma'}\, \psi_{\alpha \sigma}^\dag(0) \,
\sigma^\nu_{\sigma \sigma'} \, \psi_{\beta \sigma'}(0) \, ,
\]
are the components of the electron
spin densities in/across the leads. $\Delta_i= \mu_B g h_i$ is proportional to the magnetic field $h_i$, applied \emph{locally} to the respective impurity only, $\mu_B$ is the Bohr's magneton and $g$ is the electron gyromagnetic ratio. 
For convenience we set the Fermi velocity of the original fermions $v_F=1$. $J_{i \nu}^{\alpha \beta}$ is the set of couplings, all of which are assumed to be the parameters of the system.
 We follow [\onlinecite{SH}] and
assume $J_{i x}^{\alpha \beta} =
J_{i y}^{\alpha \beta} = J_{i \perp}^{\alpha \beta}$,
$J_{i z \pm} = (J_{i z}^{LL} \pm J_{i z}^{RR})/2$ and $J_{i z}^{LR}=J_{i z}^{RL}=0$.
The only allowed transport process is then the
spin-flip tunnelling (sometimes also called `exchange cotunnelling'), which is proportional to $J_{i \perp}^{RL}$. Inclusion of the Aharonov-Bohm phase can be done in the usual way, by supplying the spin-flip transport terms with an appropriate factor: 
We choose to do that for the tunnelling through the dot $1$, thus
\begin{eqnarray}
H_{J_{1 \perp}} = \frac{J_{1 \perp}^{RL}}{2}\left( \tau_1^+
e^{i(\lambda/2 + \varphi)} \psi_{R
  \downarrow}^\dag \psi_{L \uparrow} + \tau_1^-
  e^{i (\lambda/2 + \varphi)} \psi_{R
  \uparrow}^\dag \psi_{L \downarrow}
 \right.
 \nonumber \\
 \left.
 + \tau_1^+ e^{-i (\lambda/2 + \varphi)} \psi_{L
  \downarrow}^+ \psi_{R \uparrow} + \tau_1^- e^{-i (\lambda/2 + \varphi)} \psi_{L
  \uparrow}^\dag \psi_{R \downarrow} \right) \, .\nonumber
\end{eqnarray}
It turns out, that using the general strategy presented in Refs.~[\onlinecite{SH},\onlinecite{KondoFCS},\onlinecite{Breyel}] at the Toulouse point, when  $J_{i z -}=0$ and $J_{i z +} = 2 \pi$, the FCS could be evaluated even for $\varphi \neq 0$. Without repeating the rather lengthy but straightforward calculation we present the resulting Hamiltonian, which is a descendant of the Majorana resonant level model (MRLM): \cite{MRLM}
\begin{eqnarray} \label{oldH}
 H &=& H_0 - i J_- (b_1 + b_2) \xi_f - i J_+ (a_1 + a_2) \eta_f 
  \nonumber \\
 &-& i \sum_{i=1,2} \Delta_i a_i b_i  - i J_2 \, b_2 \xi
 \\  \nonumber
  &-& i J_1 \, b_1 \xi \cos(\lambda/2 + \varphi) 
 - i J_1 \, b_1 \eta \sin(\lambda/2 + \varphi)  \, ,
\end{eqnarray}
where $a_i$ and $b_i$ are the two different sets of local Majoranas describing the degrees of freedoms of the two dots. $\xi, \eta$ are the Majorana fields of the flavour and  $\xi_f, \eta_f$ are the Majorana fields in the spin flavour channels. The identification of the constants is as follows:  $J_{i \pm} = (J_{i \perp}^{LL} \pm J_{i \perp}^{RR})/\sqrt{2\pi a_0}$, $J_{i \perp} = J_{i \perp}^{RL}/\sqrt{2 \pi a_0}$ ($a_0$ is the lattice constant of
the underlying lattice model). 
For simplicity we have chosen the couplings $J_\pm=J_{i \pm}$ to be the same for both dots while the transversal parts are still different $J_{\perp 1,2} \to J_{1,2}$. 
The free Hamiltonian is simply
\begin{eqnarray}                      \label{H0prime}
 H_0 = i \int\, dx \, \Big[ \eta_{f}(x)
 \partial_x \eta_{f}(x) + \xi_{f}(x) \partial_x \xi_{f}(x)
 \nonumber \\
 + \eta_{}(x) \partial_x \eta_{}(x) + \xi_{}(x) \partial_x \xi_{}(x)
    ] \, .
\end{eqnarray}
The problem is quadratic in fermionic operators and thus can be solved analytically for any constellation of parameters.\cite{Breyel} An analytic expression in a closed form exists and can be found along the same lines as the CGF of the conventional noninteracting double dot setup discussed in the previous section. However, it is lengthy and therefore we relegated it to the Appendix.

\begin{figure}[htbp]
\centering
\includegraphics[width=0.5\textwidth]{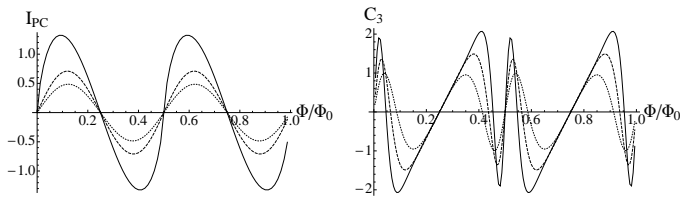}
\caption{\color{black}
\emph{Left panel:} Zero temperature PC through a double Kondo impurity ABI setup, measured in units of $\Gamma/\Phi_0$ for equal $\Gamma_{1,2}=\Gamma$ and different magnetic fields $\Delta/\Gamma = 0$, $0.5$, $1.0$ (solid, dashed, dotted lines) (we have set $\Delta_{1,2}=\Delta$).  
\emph{Right panel:} the third cumulant of the persistent current measured in units of $e^2 \Gamma {\cal T}/4\Phi_0$ for the same system in zero field $\Delta_{1,2}=0$ for different temperatures: $T/\Gamma=0.05$, $0.1$, $0.2$ (solid, dashed and dotted lines).
\label{Kondo_DD_PC_C2}}
\end{figure}

The most generic distinctive feature of the double dot Kondo ABI is the halved periodicity in $\varphi$ of all cumulants of the FCS as compared to the noninteracting double dot ABI setup, cf. Figs. \ref{Nonint_DD_IPC_C3} and \ref{Kondo_DD_PC_C2}. This is in concensus with the results of [\onlinecite{PhysRevLett.87.156803}], where the same phenomenon was predicted for the overall transmission coefficient of the double dot structure. We go considerably beyond this study and confirm this feature for the whole CGF. 

The second fundamental difference concerns the $\lambda$-periodicity of the CGF.  As was discussed in Section \ref{Generals} the CGF in the case of noninteracting systems contains incoherent $2 \pi$ periodic in $\lambda$ part as well as a coherent $4 \pi$ periodic contribution. In the present double dot Kondo case the CGF a similar effect takes place as well. CGF contains terms which are rational functions of cos and sin of $\lambda$ and $2 \lambda$. However, we tend to interpret these two contributions as a transport of single electrons and electron pairs, just like in the case of a single Kondo impurity.\cite{KondoFCS}

There two important special cases. When $\Delta_{1,2}=0$, $\Gamma_+=0$ but $\Gamma_- \neq 0$ the FCS of the single-dot system describes tunnelling of correlated electron pairs along with single-particle events.\cite{KondoFCS}  It turns out that although already at strictly $\Gamma_-=0$ not only two-particle but also single-particle processes contribute, with growing $\Gamma_-$ the admixture of the single-particle channel increases considerably leading to an overall suppression of the PC. For large $\Gamma_- \gg \Gamma$ we find the suppression factor to be of the form $\sim 1/\Gamma_-$. This kind of behaviour is to be expected since a growing contribution of a single-particle channel means a further departure from the unitary limit, at which the transmission of impurities is maximal.  

The other special case is that of finite magnetic fields $\Delta_{1,2} \neq 0$. If they become large $
|\Delta_{1,2}| \gg |\Gamma_{1,2 \perp}|$ the electron-pair processes are suppressed and the system departs from the Kondo limit. Indeed, PC degrades considerably for growing fields, see Fig. \ref{Kondo_DD_PC_C2}. 
Yet another signature of that is the decreasing asymmetry of the $I_{PC}(\varphi)$ curve with respect to the line $\varphi=\pi/4$, which is usually more pronounced for systems with high transmission. Similar phenomena can be observed for the third cumulant, see Fig. \ref{Kondo_DD_PC_C2}.

\section{Summary and conclusions}
\label{Conclusions}

To summarize, we have discussed the full counting statistics of a persistent current.
In the first part of the paper we have derived a general formula for the cumulant generating function of the charge transfer statistics in a system consisting of an open chain which is closed to a ring via a weak link. We have shown that the resulting expression for the persistent current perfectly reproduces all known results. 
We find that as a function of the counting field the CGF has a doubled periodicity as compared to the FCS of the nanostructures contacted by two independent electrodes.
We have discussed the behaviour of the second and third cumulant as functions of the magnetic flux, temperature and coupling strength to the environment. Using a slightly simplified approach we have also analyzed the FCS of a PC in a noninteracting double quantum dot Aharonov-Bohm interferometer. 

In the second part of the paper we have extended our approach to the treatment of interacting systems and applied it to a ring with an embedded Anderson impurity in a perturbatively accessible regime, finding an overall suppression of the magnitude of all cumulants.  
Furthermore, we have applied our method to transport through a double quantum dot Aharonov-Bohm interferometer in the deep Kondo limit. Here  we have made use of an integrability of the system at the Toulouse point deriving an analytical expression for the cumulant generating function in a closed form. We confirm the halved periodicity of the flux dependence in the Kondo limit, which was suggested in previous works. In agreement with the result for the FCS of a single-dot system we managed to identify single particle and electron pair processes in the corresponding CGF. 

In the future it would be particularly interesting to address the FCS of exactly solvable interacting systems in ring geometries, such as Luttinger liquids,\cite{longPC} quantum impurity models,\cite{PhysRevLett.107.206801} or consider the analytic properties of the CGF in the spirit of Ref.~[\onlinecite{PhysRevB.79.205315}]. 

}

\acknowledgements

The authors would like to thank late A. O. Gogolin, P. Schmitteckert and L.M{\"{u}}hlbacher for many fruitful discussions.
A. K. is supported by the Heisenberg programme of the DFG under grant KO 2235/5-1 and acknowledges the support by the DFG grant MU 2926/1-1.

{\color{black}
\begin{widetext}
\appendix*
\section{}
\label{App1}
Here we give an explicit expression for the CGF of the FCS for the persistent current in a double dot ABI in a Kondo limit, see Section \ref{Kondo_ABI}. To enhance readability of the formula we set $\Delta_{1}=\Delta_2=\Delta$, $\Gamma_1 = \Gamma_2 = \Gamma$ and $\Gamma_\pm=0$.
\begin{eqnarray}
 \ln \chi(\lambda) &=& 
 \frac{\cal T}{\Phi_0} \int d \omega \, 
 \ln \left\{
 8 (\omega^2 - \Delta^2)^2 \left[
  (\omega^2 - \Delta^2)^2 + 2 \omega^2 \Gamma^2 + 2 \omega^2 \Gamma^2 n_F ( 1 - n_F) (\cos(2\lambda) - 1) \right]
  \right. \nonumber \\
  &+&  \left. 8 \Gamma^2 \omega^2  (\omega^2 - \Delta^2) \left[
   (\omega^2 - \Delta^2) ( 1 + \cos \lambda \, \cos (2 \varphi) ) - i 2 \Gamma \omega (2 n_F - 1) \sin \lambda \, \sin (2 \varphi) \right]
   \right. \nonumber \\ 
   &+& \left. \Gamma^4 \omega^4 (3 - 2 \sin (2 \varphi) - 4 \cos \lambda \, \cos (2 \varphi) + \cos (2 \lambda) ) 
 \right\} \, ,
\end{eqnarray}
where the Fermi distribution function is given by $n_F = 1/(\exp(\omega/T)+1)$. 

\end{widetext}
}

\bibliography{PC_paper_v1}
\end{document}